\documentclass[11pt]{article}
\usepackage{graphicx}
\usepackage{longtable,lscape}
\usepackage{amsfonts}
\usepackage{amsmath, amsthm, amssymb}
\newcommand{\real}{{\mathbb R}}
\newtheorem*{theorem}{Theorem}
\begin{document}
\title{Experimental Evidence for Quantum Structure in Cognition}
\author{Diederik Aerts\\
        \normalsize\itshape
        Center Leo Apostel for Interdisciplinary Studies \\
         \normalsize\itshape
         Department of Mathematics and Department of Psychology\\
        \normalsize\itshape
        Vrije Universiteit Brussel, 1160 Brussels, 
       Belgium \\
        \normalsize
        E-Mail: \textsf{diraerts@vub.ac.be}
		\\ \\
		Sven Aerts\\
        \normalsize\itshape
        Center Leo Apostel for Interdisciplinary Studies \\
         \normalsize\itshape
         Department of Mathematics \\
        \normalsize\itshape
        Vrije Universiteit Brussel, 1160 Brussels, 
       Belgium \\
        \normalsize
        E-Mail: \textsf{saerts@vub.ac.be}
		\\ \\
		Liane Gabora\\
        \normalsize\itshape
        Psychology and Computer Science
         \\
         \normalsize\itshape
         University of British Columbia
         \\
        \normalsize\itshape
        Kelowna, British Columbia, Canada \\
        \normalsize
        E-Mail: \textsf{liane.gabora@ubc.ca}
		}
\date{}
\maketitle             
\begin{abstract}
\noindent
We prove a theorem that shows that a collection of experimental data of membership weights of items with respect to a pair of concepts and its conjunction cannot be modeled within a 
classical measure theoretic weight structure in case the experimental data contain the effect called overextension. Since the effect of overextension, analogue to the well-known guppy 
effect for concept combinations, is abundant in all experiments testing weights of items with respect to pairs of concepts and their conjunctions, our theorem constitutes a no-go theorem 
for classical measure structure for common data of membership weights of items with respect to concepts and their combinations. We put forward a simple geometric criterion that reveals 
the non classicality of the membership weight structure and use experimentally measured membership weights estimated by subjects in experiments from \cite{hampton1988a} to illustrate our 
geometrical criterion. The violation of the classical weight structure is similar to the violation of the well-known Bell inequalities studied in quantum mechanics, and hence suggests 
that the quantum formalism and hence the modeling by quantum membership weights, as for example in \cite{aerts2009}, can accomplish what classical membership weights cannot do.
\end{abstract}

\section{Introduction}

Many branches of mathematics, such as geometry, complexity theory, and even number theory, were originally conceived not as domains of mathematics,
 but as describing a particular domain of physical reality. It was only much later that they were conceived more abstractly, and their 
applicability to a wide range of phenomena was realized. We believe this is also proving to be the case for the mathematical formalisms
 originally developed to describe events observed in the microworld: quantum mechanics. 

Meanwhile the mathematical formalism of quantum mechanics has indeed been used successfully to model situations pertaining to domains different 
from the micro-world, for example, in economics \cite{schaden2002,baaquie2004,bagarello2006}, operations research and management
 sciences \cite{bordley1998,bordley1999}, psychology and cognition \cite{aerts1994,gabora2002a,gabora2002b,aerts2005a,aerts2005b,busemeyer2006a,busemeyer2006b,aerts2007a,aerts2007b,franco2007a,franco2007b,aerts2009}, and language and artificial intelligence \cite{aerts2004,widdow2003a,widdows2003b,vanrijsbergen2004,aerts2006,widdows2006}.

More specifically, in \cite{aerts2009} a quantum mechanical representation of experimental data corresponding to membership weights of 
items with respect to pairs of concepts and their conjunctions was elaborated. It was proven that these data cannot be modelled by a 
classical theory of membership weights, i.e. a theory where membership weights are represented within a measure theoretic structure 
(see theorems 1, 2 and 3). 

In the present paper we introduce a very simple geometrical criterion that allows the identification of the classical or non-classical nature of membership weight data gathered for pairs of concepts and their conjunctions, or more generally, collections of concepts and conjunctions of some of the pairs in these 
collections. More specifically we determine for such a collection of concepts and some of the conjunctions of these concepts a geometrical figure called a polytope (which is the higher dimensional generalization of a polygon in a real vector space) and a geometrical way of representing the measured membership weights of this collection and 
its conjunctions by means of a vector in this real vector space called a 
correlation vector. We prove that if this correlation vector is located inside the polytope a classical measure theoretic model exists for these data, while if the correlation vector is located outside of the polytope then such a model does not exist. 

\section{Membership weights on pairs of concepts and their conjunctions}
It has been shown that {\it Guppy} is neither a very typical example of {\it Pet} nor {\it Fish} but is a very typical example of {\it Pet-Fish} \cite{osherson1981}. Hence, the typicality of a specific item with respect to the conjunction of concepts can behave in an unexpected way. The problem is often referred to as the `pet-fish problem' and the effect is usually called the `guppy effect'. The guppy effect is abundant; it appears almost in every situation where concepts combine. Meanwhile many experiments and analyses of this effect and related to the problem of combining concepts have been conducted \cite{osherson1981,hampton1987,hampton1988a,hampton1988b,hampton1991,hampton1997a,hampton1997b,osherson1982,rips1995,smith1984,springer1992,storms1998}.

The guppy effect was not only identified for the typicality of items with respect to concepts and their conjunctions but also for the membership weights of items with respect to concepts and their conjunction \cite{hampton1988a}. For example, subjects rate {\it Cuckoo} a better member of the conjunction `{\it Bird and Pet}' than of the concept {\it Pet} on its own. This is a strange effect; if the conjunction of concepts behaved like a conjunction of logical propositions the second should be at least as great as the first. This deviation from what one would expect of a standard classical interpretation of conjunctions of concepts is referred to as `overextension'  \cite{hampton1988a}. Table 1 gives the list of six pairs of concepts and their conjunction for which in \cite{hampton1988a} the membership weights were measured with respect to different items, and in Table 2 the outcomes of these measurements are given for each of the items.

\section{Classical and non classical membership weights}
The behavior of a standard classical weight for a conjunction is described mathematically for the case of one pair of concepts and their conjunction in section 3 of \cite{aerts2009}. Consider weights $\mu(A_1)$, $\mu(A_2)$ and $\mu(A_1\ {\rm and}\ A_2)$ of an item $X$ with respect to a pair of concepts $A_1$ and $A_2$ and their conjunction `$A_1$ and $A_2$'. We say that they are `classical membership weights' if and only if there exists a normed measure space $(\Omega,\sigma(\Omega),P)$ and events $E_{A_1}, E_{A_2} \in \sigma(\Omega)$ of the events algebra $\sigma(\Omega)$ such that 
\begin{equation}
P(E_{A_1}) = \mu(A_1) \quad P(E_{A_2}) = \mu(A_2) \quad {\rm and} \quad P(E_{A_1} \cap E_{A_2}) = \mu(A_1\ {\rm and}\ A_2)
\end{equation}
A normed measure $P$ is a function defined on a $\sigma$-algebra $\sigma(\Omega)$ over a set $\Omega$ and taking values in the interval $[0, 1]$ such that the following properties are satisfied: (i) The empty set has measure zero, i.e. $P(\emptyset)=0$; (ii) Countable additivity or $\sigma$-additivity: if  $E_1$, $E_2$, $E_3$, $\dots$ is a countable sequence of pairwise disjoint sets in $\sigma(\Omega)$, the measure of the union of all the $E_i$ is equal to the sum of the measures of each $E_i$, i.e. $P(\bigcup_{i=1}^\infty E_i)=\sum_{i=1}^\infty P(E_i)$; (iii) The total measure is one, i.e. $P(\Omega)=1$. The triple $(\Omega,\sigma(\Omega),P)$ is called a normed measure space, and the members of $\sigma(\Omega)$ are called measurable sets. A $\sigma$-algebra over a set $\Omega$ is a nonempty collection $\sigma(\Omega)$ of subsets of $\Omega$ that is closed under complementation and countable unions of its members. Measure spaces are the most general structures devised by mathematicians and physicists to represent weights. 

We generalize this definition to the case of $n$ concepts $A_1, A_2, \ldots, A_n$ with weights $\mu(A_i)$ for each concept $A_i$, and weights $\mu(A_i\ {\rm and}\ A_j)$ for the conjunction of concepts $A_i$ and $A_j$. It is not necessary that weights are measured with respect to each one of the possible pairs of concepts. Hence, to describe this situation formally, we consider a set $S$ of pairs of indices
$S \subseteq \{(i, j)\ \vert\ i < j; i, j = 1, 2, \ldots, n\}$ corresponding to those pairs of concepts for which the weights have been measured with respect to the conjunction of these pairs. As a consequence, the following set of weights have been experimentally determined
\begin{equation}
p_i = \mu(A_i) \quad i = 1, 2, \ldots, n \nonumber \quad p_{ij} = \mu(A_i\ {\rm and}\ A_j) \quad (i, j) \in S \label{weights}
\end{equation}
We say that the set of weights in (\ref{weights}) is a `classical set of membership weights' if it has a normed measure representation, hence if there exists a normed measure space $(\Omega, \sigma(\Omega), P)$ with $E_{A_1}, E_{A_2}, \ldots, E_{A_n} \in \sigma(\Omega)$ elements of the event algebra, such that
\begin{equation}
p_i = P(E_{A_i}) \quad i = 1, 2, \ldots, n \quad
p_{ij} = P(E_{A_i} \cap E_{A_j}) \quad (i, j) \in S
\end{equation}

\section{Geometrical characterization of membership weights}
We now introduce a geometric language that makes it possible to verify the existence of a normed measure representation for the set of weights in (\ref{weights}), 
much like the characterization of Kolmogorovian probability models in \cite{pitowsky1989}. Following \cite{pitowsky1989}, we first define an $n + |S|$-tuple, called the 
$n + |S|$-dimensional correlation vector,
\begin{equation}
\overrightarrow{p} = (p_1, p_2, \ldots, p_n, \ldots, p_{ij}, \ldots)
\end{equation}
where $|S|$ is the cardinality of $S$. Denote $R(n, S) = \real^{n+|S|}$ the $n+|S|$ dimensional vector space over the real numbers. Let $\epsilon \in \{0, 1\}^n$ be an arbitrary $n$-dimensional vector consisting of $0's$ and $1's$. For each $\epsilon$ we construct the following vector $\overrightarrow{u}^\epsilon \in R(n, S)$
\begin{equation}
u^\epsilon_i = \epsilon_i \quad i = 1, 2, \ldots, n \quad
u^\epsilon_{ij} = \epsilon_i\epsilon_j \quad (i,j) \in S
\end{equation}
The set of convex linear combinations of the $u's$ is called the classical correlation polytope
\begin{equation}
c(n, S) = \{\overrightarrow{f} \in R(n, S)\ \vert\ \overrightarrow{f} = \sum_{\epsilon \in \{0, 1\}^n} \lambda_\epsilon \overrightarrow{u}^\epsilon;\ \lambda_\epsilon \ge 0;\ \sum_{\epsilon \in \{0, 1\}^n} \lambda_\epsilon = 1  \}
\end{equation}
The following theorem can now be proven similar to what was done in \cite{pitowsky1989} for the case of Kolmogorovian probabilities 
\begin{theorem}
The set of weights 
\begin{equation} \label{weights02}
p_i = \mu(A_i) \quad i = 1, 2, \ldots, n \nonumber \quad
p_{ij} = \mu(A_i\ {\rm and}\ A_j) \quad (i, j) \in S 
\end{equation}
admits a normed measure space, and hence is a classical set of membership weights, if and only if its correlation vector $\overrightarrow{p}$ belongs to the correlation polytope $c(n, S)$
\end{theorem}
Proof: Suppose that (\ref{weights02}) is a classical set of weights, and hence we have a normed measure space $(\Omega, \sigma(\Omega), P)$ and events $E_{A_i} \in \sigma(\Omega)$ such that (\ref{weights}) are satisfied. Let us show that in this case $\overrightarrow{p} \in c(n, S)$. For an arbitrary subset $X \subset \Omega$ we define $X^1=X$ and $X^0=\Omega \backslash X$. Consider $\epsilon=(\epsilon_1, \ldots, \epsilon_n) \in \{0,1\}^n$ and define $A(\epsilon)=\cap_{\epsilon}A^{\epsilon_i}_i$. Then we have that $A(\epsilon) \cap A(\epsilon')=\emptyset$ for $\epsilon\not=\epsilon'$, $\cup_{\epsilon}A(\epsilon)=\Omega$, and $\cup_{\epsilon,\epsilon_j=1}A(\epsilon)=A_j$. We put now $\lambda_\epsilon=P(A(\epsilon))$. Then we have $\lambda_\epsilon \ge 0$ and $\sum_{\epsilon}\lambda_\epsilon=1$, and $p_i=P(A_i)=\sum_{\epsilon,\epsilon_i=1}\lambda_\epsilon=\sum_{\epsilon}\lambda_\epsilon\epsilon_i$. We also have $p_{ij}=P(A_i\cap A_j)=\sum_{\epsilon,\epsilon_i=1,\epsilon_j=1}\lambda_\epsilon=\sum_{\epsilon}\lambda_\epsilon\epsilon_i\epsilon_j$. This means that $\overrightarrow{p}=\sum_{\epsilon}\lambda_\epsilon u^{\epsilon}$, which shows that $\overrightarrow{p} \in c(n, S)$. Conversely, suppose that $\overrightarrow{p} \in c(n, S)$. Then there exists numbers $\lambda_\epsilon \ge 0$ such that $\sum_{\epsilon}\lambda_\epsilon=1$ and $\overrightarrow{p}=\sum_{\epsilon}\lambda_\epsilon u^{\epsilon}$. We define $\Omega=\{0,1\}^n$ and $\sigma(\Omega)$ the power set of $\Omega$. For $X \subset \Omega$ we define then $P(X)=\sum_{\epsilon \in X}\lambda_\epsilon$. Then we choose $A_i=\{\epsilon, \epsilon_i=1\}$ which gives that $P(A_i)=\sum_{\epsilon}\lambda_\epsilon\epsilon_i=\sum_{\epsilon}\lambda_\epsilon u_i^{\epsilon}=p_i$ and $P(A_i \cap A_j)=\sum_{\epsilon}\lambda_\epsilon\epsilon_i\epsilon_j=\sum_{\epsilon}\lambda_\epsilon u_{ij}^{\epsilon}=p_{ij}$. This shows that we have a classical set of weights.

\section{The correlation polytopes for pairs of concepts and their conjunctions}

In the case of two concepts $A_1$, $A_2$ and their conjunction `$A_1\ {\rm and}\ A_2$' the set of indices is $S=\{(1,2)\}$ and the correlation polytope $c(2,S)$ is contained in the $2+|S|=3$ dimensional euclidean space, i.e. $R(2,\{1,2\})=\real^3$. Further we have four vectors $\epsilon \in \{0, 1\}^n$, namely $(0, 0), (0, 1), (1, 0)$ and $(1, 1)$, and hence the four vectors $\overrightarrow{u}^\epsilon \in \real^3$ which are the following
\begin{equation}
\overrightarrow{u}^{(0,0)}=(0, 0, 0) \quad \overrightarrow{u}^{(1,0)}=(1, 0, 0) \quad \overrightarrow{u}^{(0,1)}=(0, 1, 0) \quad \overrightarrow{u}^{(1,1)}=(1, 1, 1)
\end{equation}
This means that the correlation polytope $c(2,\{1,2\})$ is the convex region spanned by the convex combinations of the vectors $(0, 0, 0), (1, 0, 0), (0, 1, 0)$ and $(1, 1, 1)$, 
and the correlation vector is given by $\overrightarrow{p}=(\mu(A_1),\mu(A_2),\mu(A_1\ {\rm and}\ A_2))$.
It is well-known that every polytope admits two dual descriptions: one in terms of convex combinations of its vertices, and one in terms 
of the inequalities that define its boundaries \cite{aerts2007}. 
For the polytope $c(2,\{1,2\})$ the inequalities defining its boundaries are 
$0 \le \mu(A_1\ {\rm and}\ A_2)$; $\mu(A_1\ {\rm and}\ A_2) \le \mu(A_1)$; $\mu(A_1\ {\rm and}\ A_2) \le \mu(A_2)$ and $\mu(A_1)+\mu(A_2)-\mu(A_1\ {\rm and}\ A_2) \le 1$.

In Figures 1, 2 and 3 we have represented this correlation polytope $c(2,\{1,2\})$ and all the correlation vectors $\overrightarrow{p}$ for the different items (we have presented the vectors as points not to overload the figure). If the point of the correlation vector corresponding to the data of a specific item lies inside the polytope spanned by $(0, 0, 0), (1, 0, 0), (0, 1, 0)$ and $(1, 1, 1)$, it is a classical item, 
for which the membership weights can be represented within a normed measure space. 
If the point does not lie inside the polytope, the corresponding item is non-classical, 
indicating that perhaps a quantum representation, for example the one presented in \cite{aerts2009}, can be elaborated for its weights.
Since the polytope is also given by the inequalities defining its boundaries, 
points lying inside (or outside) the polytope can be characterized by their coordinates satisfying (or violating) these inequalities. 
The inequalities that define the boundaries of polytope $c(2,\{1,2\})$ are a lower dimensional variant (\cite{aerts2007} and \cite{pitowsky1989}) of the well-known
 Bell inequalities, studied in the foundations of quantum mechanics. This means that the violation of these inequalities, such as it happens by 
the data corresponding to items for which the points lie outside the polytope, has from a probabilistic perspective an analogous meaning as the violation of Bell inequalities. 
Hence these violations may indicate the presence of quantum structures in the domain where the data is collected, which makes it
plausible that a quantum model, such as for example the one proposed in \cite{aerts2009}, can be used to model the data.

\begin{table}
\caption{The list of pairs of concepts and their conjunction used in \cite{hampton1988a}}
\begin{center}
\begin{tabular}{|lll|}
\hline 
\multicolumn{1}{|l}{$A_1$} & \multicolumn{1}{l}{$A_2$} & \multicolumn{1}{l|}{$A_1$ and $A_2$} \\
\hline
{\it Furniture} & {\it Household Appliances} & {\it Furniture and Household Appliances} \\ 
{\it Food} & {\it Plant} & {\it Food and Plant} \\
{\it Weapon} & {\it Tool} & {\it Weapon and Tool} \\
{\it Building} & {\it Dwelling} & {\it Building and Dwelling} \\
{\it Machine} & {\it Vehicle} & {\it Machine and Vehicle} \\
{\it Bird} & {\it Pet} & {\it Bird and Pet} \\
\hline
\end{tabular}
\end{center}
\end{table}

\begin{table}
\small
\caption{Three of the pairs of concepts and items of experiment 4 in \cite{hampton1988a}. The non classical items are labeled by $q$ and the classical items by $c$.}
\begin{center}
\begin{tabular}{|lllll|lllll|}
\hline 
\multicolumn{1}{|l}{} & \multicolumn{1}{l}{} & \multicolumn{1}{l}{$\mu(A_1)$} & \multicolumn{1}{l}{$\mu(A_2)$} & \multicolumn{1}{l|}{$\mu(A_1{\rm and}A_2)$} & \multicolumn{1}{l}{} & \multicolumn{1}{l}{} &\multicolumn{1}{l}{$\mu(A_1)$} & \multicolumn{1}{l}{$\mu(A_2)$} & \multicolumn{1}{l|}{$\mu(A_1{\rm and}A_2)$} \\
\hline
\multicolumn{5}{|l|}{\it $A_1$=Furniture, $A_2$=Household Appliances} & \multicolumn{5}{l|}{\it $A_1$=Building, $A_2$=Dwelling} \\
\hline
{\it Filing Cabinet} & $q$ & 0.9744 & 0.3077 & 0.5263 & {\it Castle} & $c$ & 1 & 1 & 1 \\
{\it Clothes Washer} & $q$ & 0.15 & 1 & 0.725 & {\it Cave} & c & 0.2821 & 0.95 & 0.2821 \\
{\it Vacuum Cleaner} & $q$ & 0.075 & 1 & 0.3846 & {\it Phone box} & c & 0.2308 & 0.0526 & 0.02778 \\
{\it Hifi} & $q$ & 0.5789 & 0.7895 & 0.7895 & {\it Apartment Block} & $q$ & 0.9231 & 0.8718 & 0.9231 \\
{\it Heated Water Bed} & $q$ & 1 & 0.4872 & 0.775 &   {\it Library} & $q$ & 0.95 & 0.175 & 0.3077 \\ 
{\it Sewing Chest} & $q$ & 0.8718 & 0.5 & 0.55 & {\it Trailer} & $q$ & 0.35 & 1 & 0.6154 \\
{\it Floor Mat} & $q$ & 0.5641 & 0.15 & 0.2051 & {\it Jeep} & $q$ & 0 & 0.05 & 0.05 \\
{\it Coffee Table} & $q$ & 1 & 0.15 & 0.3846 & {\it Palena} & $q$ & 0.975 & 1 & 1  \\
{\it Piano} & $q$ & 0.95 & 0.1282 & 0.3333 & {\it Igloo} & $q$ & 0.875 & 1 & 0.9 \\
{\it Rug} & $q$ & 0.5897 & 0.05128 & 0.1842 & {\it Synagogue} & $c$ & 0.925 & 0.4872 & 0.4474 \\
{\it Painting} & $q$ & 0.6154 & 0.0513 & 0.1053 & {\it Tent} & $q$ & 0.5 & 0.9 & 0.55 \\
{\it Chair} & $q$ & 0.975 & 0.175 & 0.3590 & {\it Bown} & $q$ & 0.9487 & 0.8205 & 0.8974 \\
{\it Fridge} & $q$ & 0.4103 & 1 & 0.775 & {\it Theatre} & $q$ & 0.95 & 0.1282 & 0.2821 \\
{\it Desk Lamp} & $q$ & 0.725 & 0.825 & 0.825 & {\it LogCabin} & $c$ & 1 & 1 & 1 \\
{\it Cooking Stove} & $q$ & 0.3333 & 1 & 0.825 & {\it House} & $c$ & 1 & 1 & 1 \\
{\it TV} & $q$ & 0.7 & 0.9 & 0.925 & {\it Tree House} & $q$ & 0.7692 & 0.8462 & 0.85 \\
\hline
\multicolumn{5}{|l|}{\it $A$=Food, $B$=Plant} & \multicolumn{5}{l|}{\it $A$=Machine, $B$=Vehicle} \\
\hline
{\it Garlic} & $q$ & 0.9487 & 0.7105 & 0.8514 & {\it Dog Sled} & $q$ & 0.1795 & 0.925 & 0.275 \\
{\it Toadstool} & $q$ & 0.1429 & 0.6061 & 0.2727 & {\it Dishwasher} & $q$ & 1 & 0.025 & 0 \\
{\it Steak} & $c$ & 1 & 0 & 0 & {\it Backpack} & $c$ & 0 & 0 & 0 \\
{\it Peppercorn} & $q$ & 0.875 & 0.6207 & 0.7586 & {\it Bicycle} & $q$ & 0.85 & 0.975 & 0.95 \\
{\it Potato} & $q$ & 1 & 0.7436 & 0.9 & {\it Sailboat} & $c$ & 0.5641 & 0.8 & 0.4211 \\
{\it Raisin} & $q$ & 1 & 0.3846 & 0.775 & {\it Roadroller} & $q$ & 0.9375 & 0.9063 & 0.9091 \\
{\it Mint} & $q$ & 0.8718 & 0.8056 & 0.8974 & {\it Raft} & $c$ & 0.2051 & 0.725 & 0.2 \\
{\it Sunflower} & $q$ & 0.7692 & 1 & 0.775 & {\it Elevator} & $q$ & 0.9744 & 0.6 & 0.7949 \\
{\it Seaweed} & $q$ & 0.825 & 0.9744 & 0.8684 & {\it Course liner} & $q$ & 0.875 & 0.875 & 0.95  \\
{\it Sponge} & $q$ & 0.0263 & 0.3421 & 0.0882 & {\it Automobile} & $c$ & 1 & 1 & 1  \\
{\it Bread} & $q$ & 1 & 0.0769 & 0.2051 & {\it Horsecart} & $q$ & 0.3846 & 0.95 & 0.2895 \\
{\it Cabbage} & $q$ & 1 & 0.9 & 1 & {\it Skateboard} & $q$ & 0.2821 & 0.8421 & 0.3421 \\
{\it Eucalyptus} & $q$ & 0.1622 & 0.8974 & 0.3243 & {\it Bus} & $c$ & 1 & 1 & 1 \\
{\it Poppy} & $q$ & 0.3784 & 0.8947 & 0.5405 & {\it Bulldozer} & $q$ & 1 & 0.925 & 0.95 \\
{\it Mushroom} & $q$ & 1 & 0.6667 & 0.9 & {\it Lawn-mower} & $q$ & 0.975 & 0.1053 & 0.2632 \\
{\it Lettuce} & $q$ & 1 & 0.925 & 1 & {\it Ski Lift} & $q$ & 1 & 0.5897 & 0.875 \\
\hline
\multicolumn{5}{|l|}{\it $A$=Weapon, $B$=Tool} & \multicolumn{5}{l|}{\it $A$=Bird, $B$=Pet}\\
\hline
{\it Ruler} & $q$ & 0.05 & 0.9 & 0.1538 & {\it Dog} & $c$ & 0 & 1 & 0 \\
{\it Toothbrush} & $c$ & 0 & 0.55 & 0 & {\it Cuckoo} & $q$ & 1 & 0.575 & 0.8421 \\
{\it Chisel} & $q$ & 0.4 & 0.975 & 0.6410 & {\it Parakeet} & $c$ & 1 & 1 & 1 \\
{\it Axe} & $q$ & 0.875 & 1 & 0.975 & {\it Cat} & $c$ & 0 & 1 & 0\\
{\it Screwdriver} & $q$ & 0.3 & 1 & 0.625 & {\it Lark} & $q$ & 1 & 0.275 & 0.4872 \\
{\it Arrow} & $q$ & 1 & 0.225 & 0.575 & {\it Heron} & $q$ & 0.9412 & 0.1515 & 0.2581 \\
{\it Knife} & $c$ & 1 & 0.975 & 0.975 & {\it Peacock} & $q$ & 1 & 0.4 & 0.5789 \\
{\it Rifle} & $q$ & 1 & 0.35 & 0.5 & {\it Cow} & $q$ & 0 & 0.425 & 0.025 \\
{\it Whip} & $q$ & 0.875 & 0.2632 & 0.625 & {\it Toucan} & $q$ & 1 & 0.6154 & 0.8026 \\
{\it Hammer} & $q$ & 0.575 & 1 & 0.8 & {\it Parrot} & $c$ & 1 & 1 & 1 \\
{\it Scissors} & $q$ & 0.6053 & 0.9744 & 0.7692 & {\it Mynah Bird} & $q$ & 1 & 0.8710 & 0.8438 \\
{\it Spoon} & $q$ & 0 & 0.752 & 0.075 & {\it Raven} & $q$ & 1 & 0.2368 & 0.4 \\
{\it Spear} & $q$ & 1 & 0.275 & 0.7179 & {\it Elephant} & $c$ & 0 & 0.25 & 0 \\
{\it Chain-saw} & $q$ & 0.55 & 1 & 0.75 & {\it Goldfish} & $c$ & 0 & 1 & 0 \\
{\it Club} & $q$ & 1 & 0.3590 & 0.775 & {\it Homing Pigeon} & $q$ & 1 & 0.775 & 0.8974 \\
{\it Razor} & $q$ & 0.625 & 0.775 & 0.825 & {\it Canary} & $c$ & 1 & 1 & 1 \\
\hline 
\end{tabular}
\end{center}
\end{table}

\begin{figure}[htp]
\centerline {\includegraphics[width=14cm]{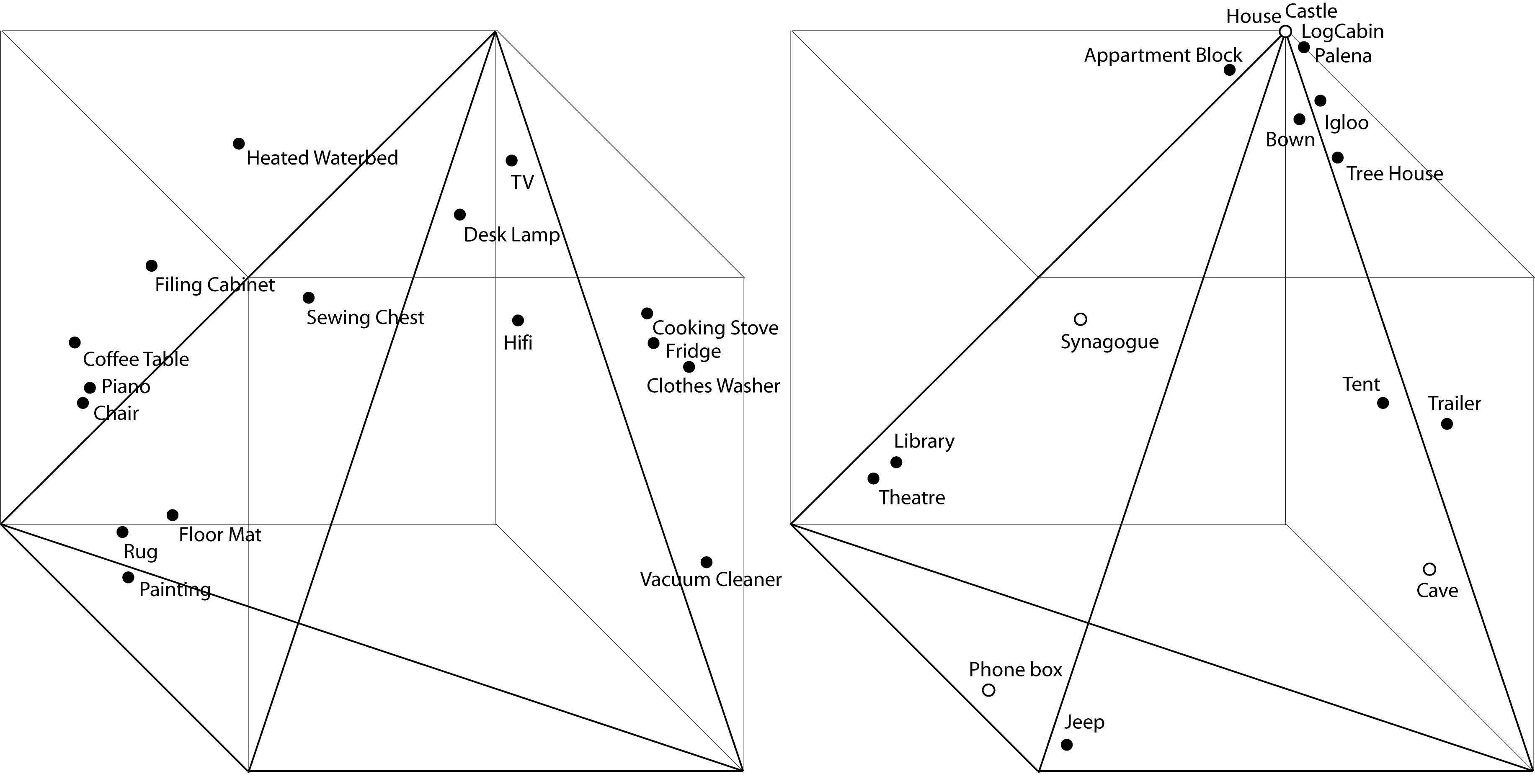}}
\caption{The polytopes for the concepts {\it Furniture} and {\it Household Appliances} and the concepts {\it Building} and {\it Dwelling}. The classical items are {\it Castle}, {\it Cave}, {\it Phone Box}, {\it Synagogue}, {\it Log Cabin} and {\it House}. The other items are non classical.}
\end{figure}

\begin{figure}[htp]
\centerline {\includegraphics[width=14cm]{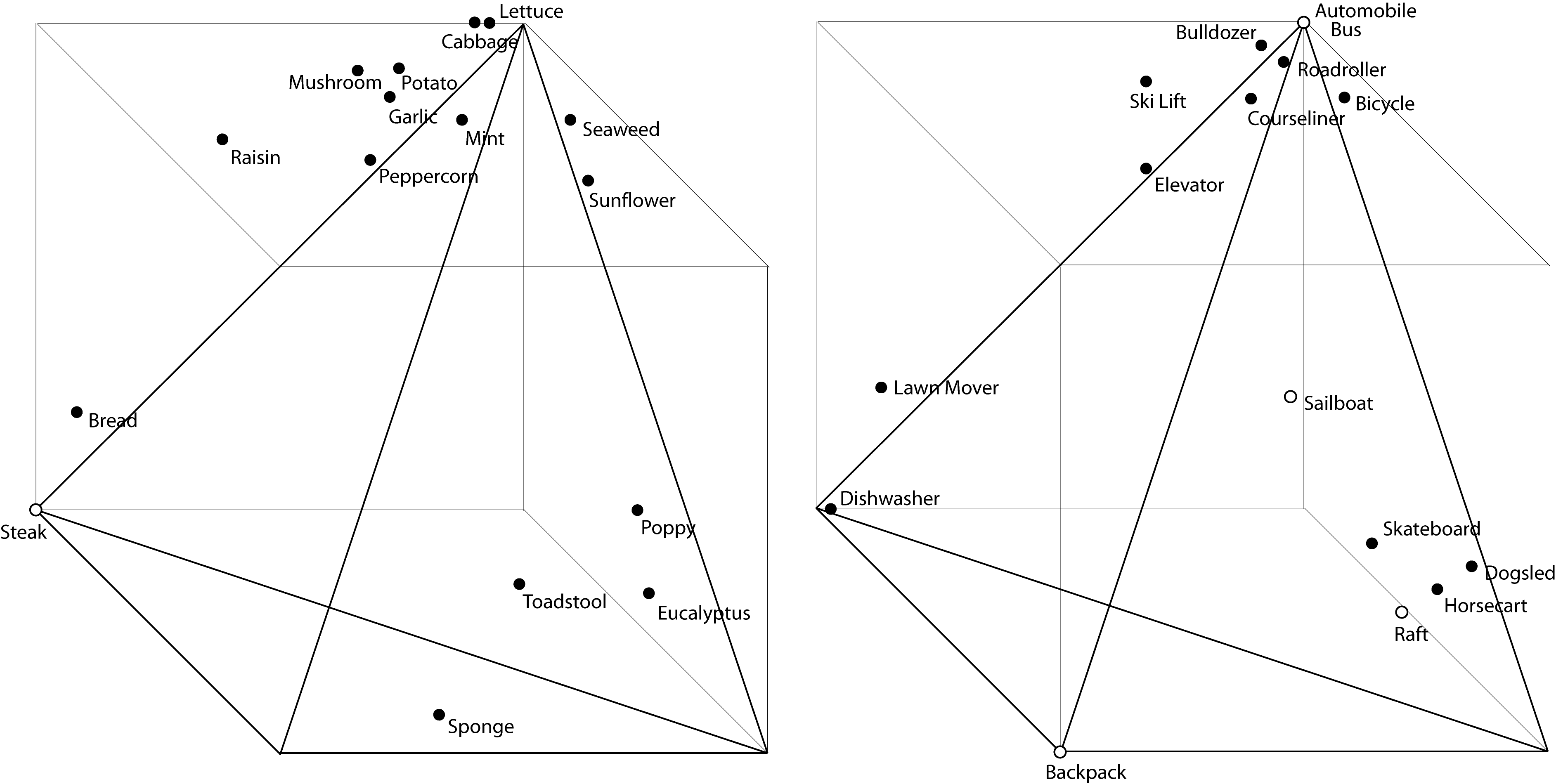}}
\caption{The polytopes for the concepts {\it Food} and {\it Plant} and the concepts {\it Machine} and {\it Vehicle}. The classical item are {\it Steak}, {\it Backpack}, {\it Automobile}, {\it Bus}, {\it Sailboat} and {\it Raft}. The other items are non classical.}
\end{figure}

\begin{figure}[htp]
\centerline {\includegraphics[width=14cm]{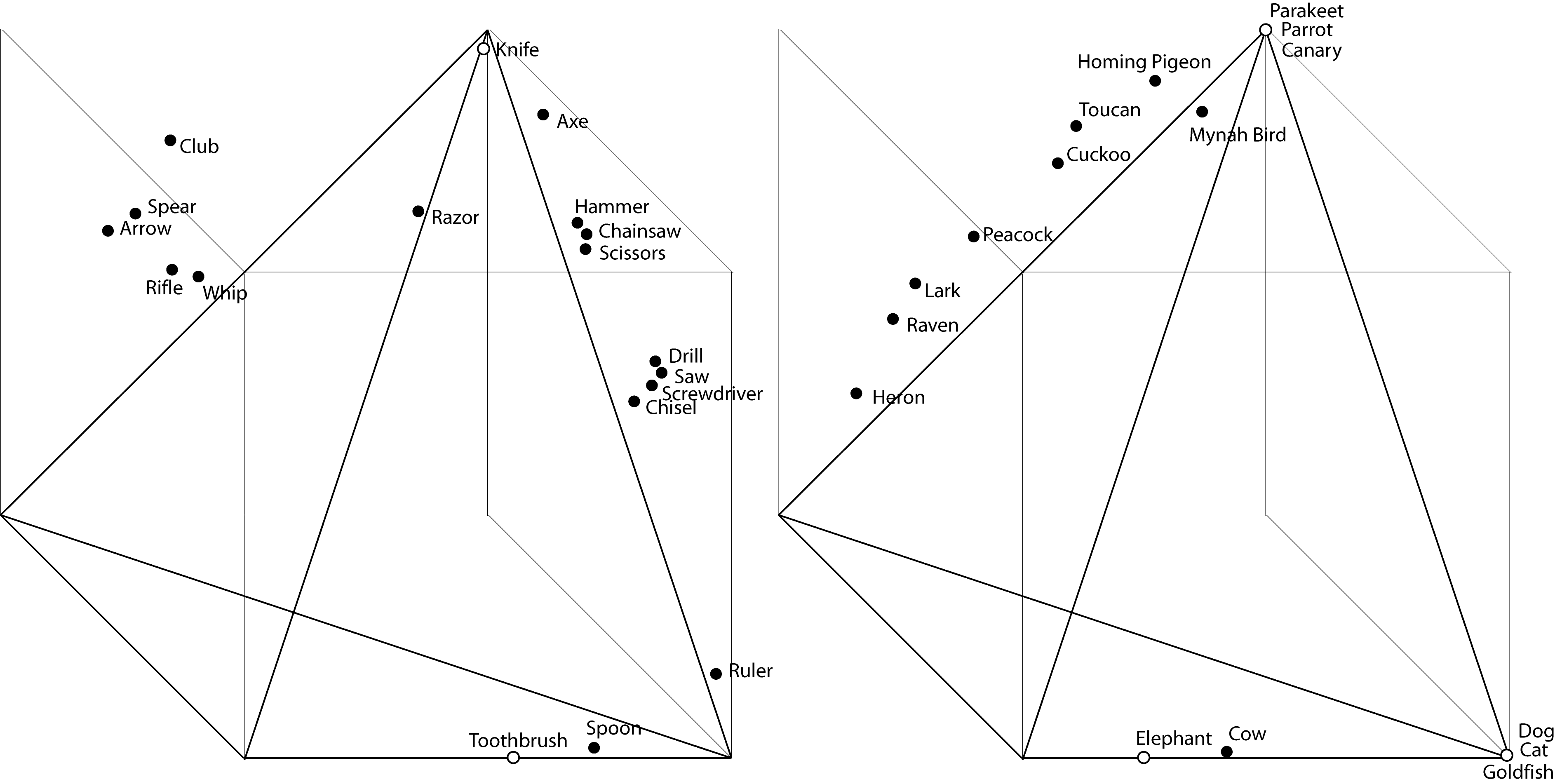}}
\caption{The polytopes for the concepts {\it Weapon} and {\it Tool} and the concepts {\it Bird} and {\it Pet}. The classical item are {\it Knife}, {\it Toothbrush}, {\it Elephant}, {\it Dog}, {\it Cat}, {\it Goldfish}, {\it Parakeet}, {\it Parrot} and {\it Canary}. The other items are non classical.}
\end{figure}

\section{Conclusion}
If two concepts are combined to form a conjunction we can measure the membership weights of items with respect to each of these concepts and also with respect to their conjunction. 
If the `conjunction of concepts' behaved like a classical logical conjunction of propositions does, we would expect that the membership weight of an item with respect to the conjunction 
would never be bigger than the membership weight of this item with respect to one of the concepts. Experiments show that this is not the case, and this counterintuitive effect is referred to as both the 
guppy effect \cite{osherson1981} and overextension \cite{hampton1988a} in the literature. It has been shown elsewhere that a quantum description can model this overextension while classical measure theoretic 
structures cannot \cite{aerts2009}. In this paper we have elaborated a simple geometric method to identify the membership weights of items with respect to the conjunction of concepts that 
cannot be modeled within a classical measure theoretic structure. We do this for the general situation of a set of $n$ concepts and a set of conjunctions between these concepts. The method 
consists of determining a convex polytope, and making each of the items correspond to a correlation vector in the real vector space where also the polytope is defined. We prove that if the 
correlation vector is contained in the polytope, the considered set of membership weights can be modeled within a classical measure space, while if the correlation vector is not contained in 
the polytope it cannot. We apply this geometrical characterization method to the set of data collected in \cite{hampton1988a} and see that most of the tested items have membership weights for 
which the correlation vector falls outside of the polytope, and hence these membership weights cannot be modeled within a classical measure space (see Figures 1, 2 and 3).

The experimental data for which we show in the present article by means of the polytope criterion that they are non classical, i.e. cannot be represented within a classical measure structure, are all `conjunction data', meaning that they are membership weights of items with respect to the conjunction of two concepts. The phenomenon of structural non classicality that we put into evidence in this article is however much more general and does not only appear with membership weights of conjunctions. For example, it appears in a very analogous way for disjunctions of concepts. as experimentally shown in \cite{hampton1988b} and theoretically analyzed in \cite{aerts2009}, theorems 4, 5 and 6. Following our contextual theory of concepts developed in \cite{aerts2005a,aerts2005b}, 
we have good reason to believe that the effect appears whenever concepts are combined. 
Unfortunately, the non-classical effect is difficult to identify for an arbitrary combination of concepts, since we do not have a simple mathematical characterization, (like we have for conjunction and disjunction) of what the classical structure of such an arbitrary combination would be. 
In \cite{hampton1988b} next to the disjunction also the negation is investigated, and also there significant deviations of what one would expect classically from the logical structure of a negation are measured. We have not yet analyzed these effects on the negation with respect to the quantum models developed in \cite{aerts2007a,aerts2007b,aerts2009}, but plan to do so in future work. It would be interesting to make this type of experiments with the remaining not yet tested simple logical connective which is called `the implication'. The appearance of this type of non-classical weights is however not limited to the domain of concepts and their combinations. In decision theory and in economics different types of situations have been studied entailing a very similar type of non classical weight structure than the one we consider in the present paper \cite{aertsdhooghe2009}. The disjunction effect \cite{tversky1992} and the conjunction fallacy \cite{tversky1982,tversky1983} are the best known ones, and the disjunction effect has been modeled by a quantum mechanical description in \cite{busemeyer2006b}, while the conjunction fallacy was analyzed with respect to quantum mechanical modeling in \cite{franco2007a}. In \cite{bar-hillel1993} a `disjunction fallacy' is experimentally identified, by considering disjunctions that are not combinations of concepts but already received a name of themselves, such as for example {\it Natural Sciences} being the disjunction of {\it Astronomy}, {\it Physics}, {\it Chemistry}, {\it Biology} and {\it Earth Sciences}. The deviation from how classically this type of disjunction should behave is shown in \cite{bar-hillel1993} to be very big. 
Hence this demonstrates that the non classicality does not find its origin in a kind of `wrong application of the combination rules', a hypothesis put forward in \cite{gavanski1991}. 
It shows that the effect is also present for disjunctions and conjunctions that are single concepts and not combinations. 
From the perspective of the quantum model developed in \cite{aerts2007a,aerts2007b,aerts2009} we have put forward an explanation for the non classicality,
 due to the fact that for combinations of concepts equally the single new emergent concept as well as the combination as a logical connective play a role in the influence provoked by the 
 conceptual landscape surrounding the decision situation \cite{aertsdhooghe2009}. 
A similar simple criterion with polytopes can be worked out for situations of non classicality of other combinations of concepts, 
and also for the non classical phenomena identified in decision theory, such as the disjunction effect and the conjunction fallacy. 
We plan to elaborate this in future work.

We note that have formulated all hypothesis and claims in the present article by considering the notion of a `normed measure space'
 and its elements representing the membership weights of the items with respect to concepts. 
Alternatively, we could equally well have considered `the probability for a specific subject to choose in favor of membership' 
in replacement of `the membership weight' as central element. If we would have done so our theorem would become a theorem 
on probability models instead of a theorem on unitary measure spaces, and it would be mathematically completely equivalent 
with Pitowsky's main theorem in \cite{pitowsky1989}.

 The geometrical identification presented here gives rise to a demarcation similar to the violation of the well-known Bell inequalities in physics, which is generally regarded as experimental evidence 
for the need of a fundamental change in the classical paradigm to describe the process under consideration. 
Hence, Hampton's membership weight data giving rise to a situation equivalent to the violation of Bell inequalities, constitutes a strong argument in favor of the fact that quantum 
structure would be at work within the mechanism giving rise to these data, hence within human cognition.
If so, then these experiments constitute a pioneering example of experimentally tested quantum structure in cognition
performed by a psychologist in tempore non suspecto. It also would mean that only a non classical description, for example one based on quantum mechanics eventually as the one elaborated 
in \cite{aerts2009}, is able to model the mechanism giving rise to the data.

\section{Acknowledgments}

This work was supported by grant G.040508N from the FWO-Flanders, Belgium, and a grant from the Social Sciences and Humanities Research Council of Canada.

\end{document}